\documentclass
[preprint,prl,showpacs,noshowkeys,10pt,twocolumn,tightenlines]{revtex4}%
\usepackage{amsfonts}
\usepackage{amsmath}
\usepackage{amssymb}
\usepackage[dvips]{graphicx}%
\begin{document}
\preprint{ }
\title{Conductance relaxation in the Electron-Glass; Microwaves versus infra-red response}
\author{Z. Ovadyahu}
\affiliation{Racah Institute of Physics, The Hebrew University, Jerusalem 91904, Israel }

\begin{abstract}
We study the time-dependent conductance of electron-glasses excited by
electromagnetic radiation at microwaves (MW) and infra-red frequencies. In
either case the conductance G is enhanced during exposure but its time
dependence after the radiation is turned off is qualitatively different
depending on the frequency. For comparison, results of excitation produced by
a gate-voltage and temperature changes are also shown. The glassy nature of
the system allows us to demonstrate that the MW-enhanced conductance is
\textit{not} due to heating. These findings are discussed in terms of an
energy E$_{c}$ that characterizes the equilibrium charge distribution of the electron-glass.

\end{abstract}
\maketitle

One of the characteristic properties of the electron-glass \cite{1} is a slow
relaxation of an out-of-equilibrium conductance. At equilibrium the
conductance G is at a local minimum, and any agent that takes the system out
of equilibrium gives rise to an excess conductance \cite{2}. In most
electron-glasses studied to date, the preferred method of exciting the system
was a sudden change of the carrier-concentration using a gate \cite{2,3}. This
usually led to a logarithmic relaxation\ of the excess conductance, which was
theoretically accounted for by several authors \cite{4}. Being an electronic
glass, there are many other ways to excite the system and follow its
relaxation dynamics.

In this note we study the effects of exposing electron-glasses to
electromagnetic radiation at microwaves (MW) frequencies and compare the
results with the response obtained at infra-red (IR) frequencies. When either
type of irradiation is turned on G promptly increases. However, the excess
conductance $\Delta$G caused by the radiation decays in a different way upon
turning it off; $\Delta$G following exposure to IR exhibits sluggish
relaxation that may last several hours while the MW enhanced G disappears
rather quickly. This qualitative difference between the IR and MW excitations
is interpreted as evidence for a characteristic energy E$_{c},~$presumably,
the Coulomb interaction associated with the equilibrium distribution of the
charge carriers among the (localized) electronic states.

Thirty two samples were measured in this study. These were thin films of
either crystalline or amorphous indium-oxide (In$_{2}$O$_{3-x}$ and In$_{x}$O
respectively) prepared on 110$\mu$m glass substrates with a metallic electrode
deposited on their backside to act as gate. Lateral size of the samples used
here were 0.2-1mm. Their thickness (typically, 30-50\AA ~for In$_{2}$O$_{3-x}$
and 80-200\AA ~for In$_{x}$O) and stoichiometry were chosen such that at the
measurement temperatures (at or close to 4K), samples had sheet resistance
R$_{\square}$ in the range 3M$\Omega$-3G$\Omega$. Fuller details of sample
preparation, characterization, and measurements techniques are given elsewhere
\cite{5,6}.

Several sources were used for MW excitation; Gunn diodes (up to 100GHz),
Klystrons (10 and 35GHz), and a high power synthesizer (HP8360B, 2-20GHz, and
up to $\approx$320mW$\equiv$25dBm output-power). The data shown here are
mostly based on output of the latter source being fed to the sample chamber
via a coaxial cable. The MW power at the sample stage was measured to be
linear with the synthesizer output-power. A LED diode (operating
at$\approx2\cdot10^{14}\sec^{-1}$), placed$\approx$2mm from the sample, was
used to generate the IR.

The response of typical In$_{2}$O$_{3-x}$ and In$_{x}$O samples to MW
radiation is compared with the response to IR radiation in Fig.~1. Also shown
is the effect of quickly changing the gate voltage V$_{g}$ which has been the
most common method to excite the electron-glass \cite{3}. Here we wish to
focus on a systematic \textit{qualitative} difference between the effect
produced by IR excitation (or changing V$_{g}$), and that of the MW
illumination; namely, the lack of slow relaxation in $\Delta$G(t) caused by MW
radiation. The absence of a long relaxation tail following excitation by MW
was confirmed throughout the range 0.9-100GHz using different sources.
Excitation by visible light sources, on the other hand, led to the same
qualitative behavior as observed using the IR excitation.%
\begin{figure}
[ptb]
\begin{center}
\includegraphics[
trim=0.000000in 1.543609in 0.000000in 1.085739in,
height=3.6512in,
width=3.3243in
]%
{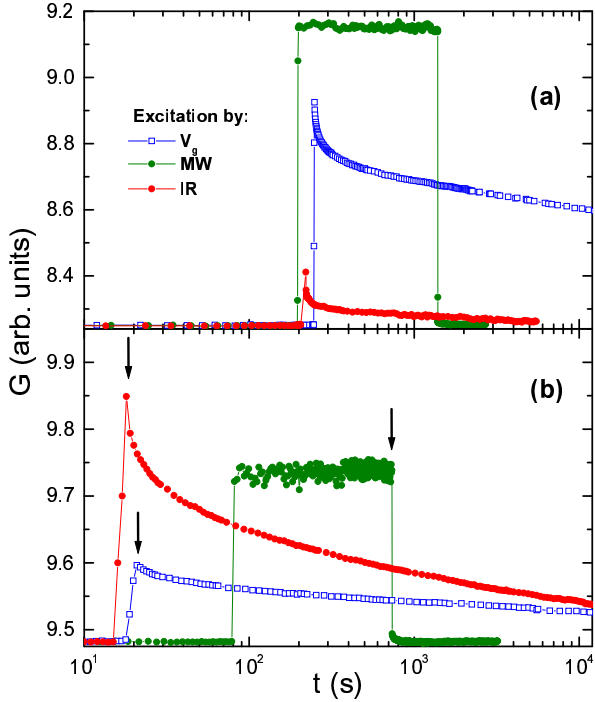}%
\caption{Time dependence of the conductance of typical electron glasses under
three excitation protocols, each starting from equilibrium; V$_{g}$ - the gate
to sample voltage is swept from 0 to 105V with rate=25V/s; IR - 60$\mu$W
source operated for 2 seconds; MW-frequency of 2.633GHz and source power of
21dBm held for few hundred seconds then turned off. Samples are: (a) In$_{2}%
$O$_{3-x}$ thickness 34\AA ~R$_{\square}$=12M$\Omega$ n=4.4$\cdot$10$^{19}%
~$cm$^{-3}.$ (b) In$_{x}$O thickness 100\AA ~R$_{\square}$=5M$\Omega$
n=9.05$\cdot10^{19}$ cm$^{-3}.$ Arrows mark the onset of the respective
relaxation process.}%
\end{center}
\end{figure}

As the main difference between the IR (and visible) and the MW fields is their
frequency $\omega,$ the different behavior of $\Delta$G(t) following
excitation by these agents may be a hint on the existence of a characteristic
energy E$_{c}$ such that E$_{c}\eqslantgtr\hbar\omega$ for the applied MW and
E$_{c}\eqslantless\hbar\omega$ for the IR illumination. A natural candidate
for E$_{c}$ is the Coulomb energy associated with the inter-electron
interaction which is of order $e^{2}n^{-3}/\kappa$ where n is the carrier
concentration and $\kappa$ is the dielectric constant. It was recently shown
that $e^{2}n^{-3}/\kappa$ is of the same order as the width of the "memory
dip" which is the characteristic signature of the (intrinsic) electron-glass
\cite{6}. For the indium-oxides this energy spans the range 6-80meV depending
on the carrier-concentration n. The highest energy of the MW used here
($\approx$0.42meV) and the energy of the IR source ($\approx$830meV), straddle
this energy range.

The proposed picture is as follows. The system is in equilibrium when the
occupation of the electronic states, under a given potential landscape and
temperature, minimizes the free energy. In the interacting system this leads
to a specific organization of the way the localized states are occupied. The
Coulomb interaction introduces correlations between states occupation and
their spatial coordinates, a process that, among other things, leads to a
Coulomb gap in the single-particle density of states \cite{7}. Formation of
this configuration from an excited state (where the sites are randomly
distributed) is a slow process for several reasons: The lowest energy state of
the system cannot be reached without the (inherently slow) many-particle
dynamics being involved \cite{8}. Dynamics is further impeded by hierarchical
constraints \cite{9} imposed by the Coulomb interactions, and most
importantly, quantum friction, which is appreciable in systems with large
carrier-concentration is the dominant factor in bringing about sluggish
relaxation \cite{6,10}. A substantial change of such a configuration requires
an energy investment of $\eqslantgtr$E$_{c},~$which is of the order of the
Coulomb energy. Only excitation agents that can impart an
energy-quantum$~\eqslantgtr$E$_{c}~$may randomize this configuration. Since
re-establishing an equilibrium configuration is a sluggish process, $\Delta
$G(t) produced by such an excitation agent will decay slowly. This is the case
for excitation by IR as well as due to a sudden change in V$_{g}$. By
contrast, the conductance of the system may be increased without randomizing
the equilibrium configuration, e.g., when extra energy is imparted to the
hopping process (\textit{from a source with energy quanta smaller than }%
E$_{c}$),$~$in which case the relaxation could be fast. This is presumably the
case with the MW excitation (a possible specific scenario is mentioned in the
summary below). In general, injecting energy into the system will enhance the
conductance rather immediately in any case. Therefore, the
excitation-relaxation process will appear asymmetric in $\Delta$G(t) plots (as
in Fig.~1 for the gate and IR excitation), and symmetric (for MW excitation).

When an applied ac field increases G in a system with a negative temperature
coefficient of resistance, it is natural to assume that \textit{heating} is
the cause, as was conjectured by Ben-Chorin et al \cite{2}. For
strongly-localized samples it takes a rather small change of temperature
$\Delta$T to get an appreciable $\Delta$G/G$.$ Fig.~2 illustrates this point.
The figure shows how $\Delta$G/G depends on the MW power for several samples,
and the inset delineates the temperature change needed to affect such $\Delta
$G/G. This plot shows that the required $\Delta$T is less than 8\% of the bath
temperature even for the two samples with the highest MW sensitivity. However,
although some heating accompanies the MW radiation (and, in fact, the IR
radiation as well), it turns out to be much less than this estimate, and the
bulk of the MW-induced $\Delta$G is \textit{not} a result of the sample being
heated-up by the MW radiation.%
\begin{figure}
[ptb]
\begin{center}
\includegraphics[
trim=0.000000in 3.013551in 0.000000in 1.207007in,
height=2.9879in,
width=3.3243in
]%
{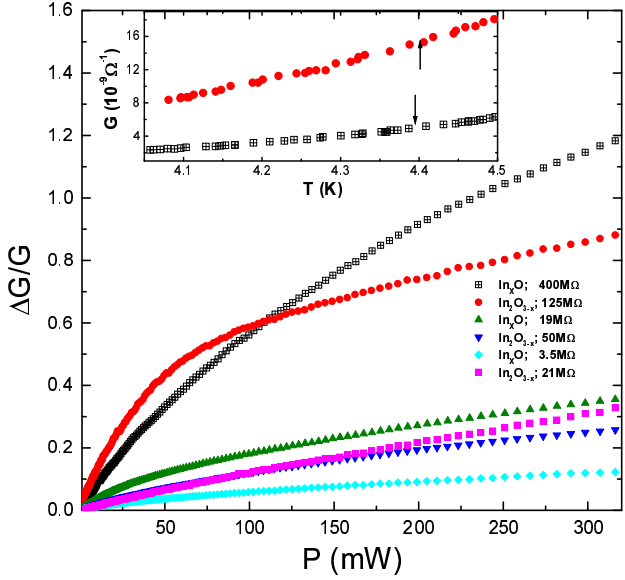}%
\caption{Relative conductance change versus MW power for six of the samples
that showed the highest sensitivity to MW. Significantly, n in these samples
was smaller than 10$^{20}$ cm$^{-3}$; $\Delta$G/G for samples with larger n
never exceeded 5\% at full MW power. Inset shows G(T) for the two samples with
the highest sensitivities. Arrows mark the T at which G attains the MW-induced
conductance (at full power).}%
\end{center}
\end{figure}

That sample heating cannot explain the MW induced $\Delta$G could have been
already deduced from the data in Fig.~2; The sublinear $\Delta$G versus
MW-power, in a range where both G and the power dissipated to the bath should
be linear over $\Delta$T$~$(given $\Delta$T/T$\ll$1), is incompatible with
heating mechanism. In the following we describe experiments that utilize the
unique transport features of the electron-glass to get a more direct test of
this issue.

The first experiment, described in Fig.~3 employs the memory-dip (MD) as a
thermometer. The MD is the characteristic feature in G(V$_{g}$) centered at
the gate voltage where the system has equilibrated, and its relative amplitude
is extremely sensitive to temperature \cite{11,5}. The figure clearly
demonstrates that MW radiation, while increasing the conductance by a certain
$\Delta$G, it leaves the MD amplitude essentially unchanged (Fig.~3a). To get
the same $\Delta$G by raising the bath temperature one has to use $\Delta$T
that produces a distinct change in the magnitude of the MD (Fig.~3b).%
\begin{figure}
[ptb]
\begin{center}
\includegraphics[
trim=0.000000in 0.723071in 0.000000in 0.241402in,
height=4.3457in,
width=3.3243in
]%
{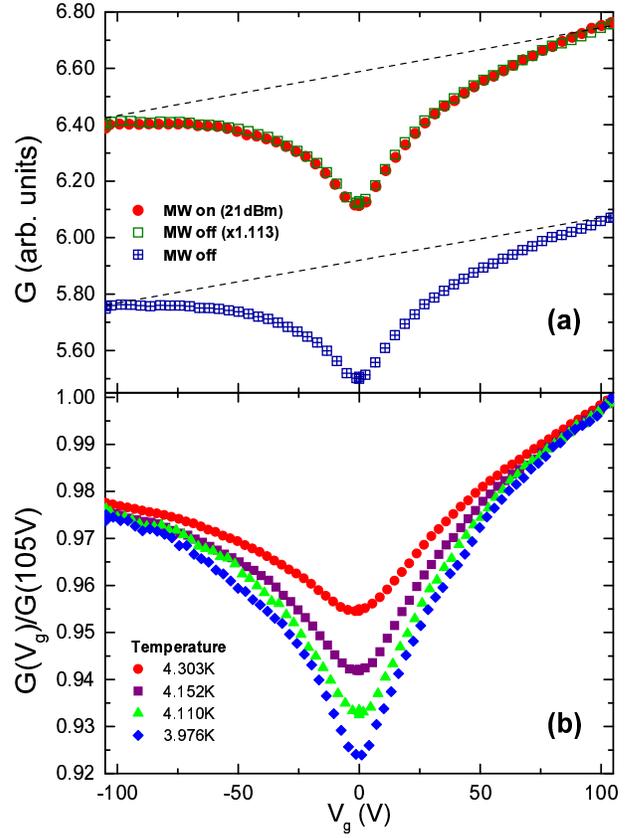}%
\caption{G(V$_{g}$) traces taken under various conditions for a typical
In$_{2}$O$_{3-x}$ sample (thickness 34\AA ~R$_{\square}$=28M$\Omega$
n=4.3$\cdot$10$^{19}~$cm$^{-3}$). Note the memory-dip centered at V$_{g}$=0
where the sample was first equilibrated for 46 hours. (a) With and without MW
radiation (at f=5.487GHz) at a bath temperature T=4.11K. (b) At different
temperatures (with MW off and equilibration period of 2 hours at each T). Note
that the slopes of the anti-symmetric field-effect (see, \cite{10} and
references therein), depicted by dashed lines, are essentially unaffected by
the MW. This suggests that the enhanced G is not due to an increase in
carrier-concentration. }%
\end{center}
\end{figure}
Obviously, changing G by $\Delta$T yields a different physical situation than
the effect of changing it by MW radiation. This can also be seen from another
angle as described in the experiment shown in Fig.~4. Here, each excitation
agent is applied starting from equilibrium, and is maintained for
$\approx750~\sec.$ It is then turned off and the ensuing behavior of $\Delta
$G(t) is recorded. This protocol is illustrated for the $\Delta$T
excitation-run in Fig.~4a (the respective protocol for the MW excitation can
be seen e.g., in Fig.~1). The time dependence of $\Delta$G(t), after
conditions are set at \textit{status-quo ante, }is shown in 4b allowing a
comparison between the two protocols. Note that after the excitations are
turned off, at t=t$^{\ast}$, G(t%
$>$%
t$^{\ast}$) relaxes back to its equilibrium value for both protocols, but in a
different way: the amplitude of $\Delta$G(t%
$>$%
t$^{\ast}$) for the MW protocol is relatively small, and it relaxes rather
fast while a glassy (logarithmic) relaxation is observed in the $\Delta$T
protocol and a measurable $\Delta$G(t%
$>$%
t$^{\ast}$) persists for a much longer time. On the basis of these experiments
we estimate that the temperature increment the samples gain from the radiation
is smaller by 30-50 than the $\Delta$T that would increase G by the MW induced
value.
\begin{figure}
[ptbptb]
\begin{center}
\includegraphics[
trim=0.000000in 0.843205in 0.000000in 0.243668in,
height=4.2938in,
width=3.3243in
]%
{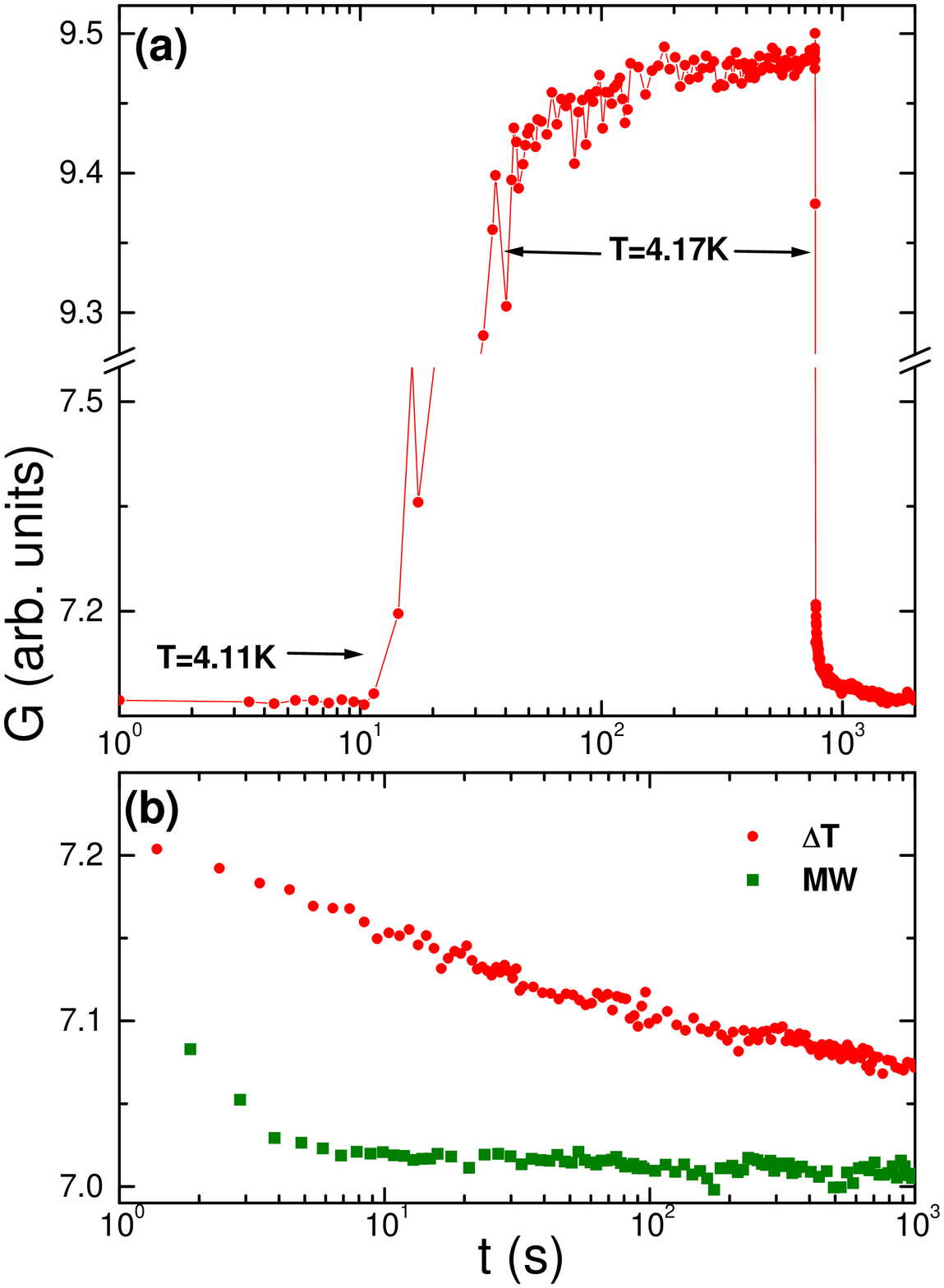}%
\caption{Comparison between the temporal dependence of G subjected to $\Delta
$T and MW protocols (see text). Sample is a In$_{2}$O$_{3-x}$ (thickness
3{4\AA  }~R$_{\square}$ =28M$\Omega$ n=4.27\textperiodcentered10$^{19}$
cm$^{-3}$). (a) Illustrating the protocol for $\Delta$T=60mK chosen to match
the $\Delta$G/G$\approx$30\% produced by exposing the sample to MW (power of
25dBm at f=2.416GHz) in the MW protocol. (b) The time dependence of the excess
conductance of the two protocols.}%
\end{center}
\end{figure}

These experiments tell us also what \textit{is} the difference between
exposing the system to MW radiation and raising its temperature by $\Delta$T;
The latter presumably affects the equilibrium configuration while the MW does not.

A question arises here; As $\Delta$T$\ll\frac{E_{c}}{k_{B}},$ how does one
reconcile the respective $\Delta$G(t) behavior with our conjecture that it
takes an energy-quantum $\eqslantgtr$E$_{c}$ to modify the equilibrium
configuration to get glassy relaxation? The answer is that, in contrast with
MW radiation that has a high frequency cutoff, at a temperature T the system
still experiences phonons with $\hbar\omega\gg k_{B}$T, albeit with
exponentially diminished probability. Letting $\Delta$T operate on the system
for a finite time, affects the equilibrium configuration through the presence
of thermal phonons with energies $\eqslantgtr$E$_{c}$. The exponential
sensitivity of the MD amplitude to temperature apparent e.g., in Fig.~3a (see
also \cite{6,11}), is one manifestation of this effect.

Applying a non-Ohmic longitudinal field seems to have a similar effect as
raising the temperature \cite{12} in that it leads to slow relaxation. This
may be partly due to real heating \cite{13} or to\ a field-created new current
paths with the accompanying re-organization of the equilibrium configuration
of the occupied states.

In summary, we investigated the behavior of the excess conductance created by
electromagnetic radiation applied to several electron-glasses. A systematic
qualitative difference is found between sources depending on the
quantum-energy of the associated photon. This is argued to be consistent with
a characteristic Coulomb energy relevant for the electron-glass. Our
conjecture is amenable to a more refined test by carrying out optical
excitation experiments, in particular, over the energy range 6-80meV which is
the range of energies associated with the glassy phase of the indium-oxides.
This should be possible using Synchrotron radiation with a series of samples
having different carrier-concentration n. We expect a crossover from MW-like
to IR-like response around a frequency $\omega$ that scales with the MD width
(that, in turn, is determined by the carrier-concentration \cite{14,10}).
Electron-glasses with low n might have offered a more convenient frequency
range; A crossover frequency $\omega_{c}$ of 0.4-0.7THz was reported in
G($\omega$) measurements on Si:P samples \cite{15} with n in the range
1.6$\cdot10^{18}-$3$\cdot10^{18}$cm$^{-3}$,$~$ smaller by 1-3 orders of
magnitude than the carrier-concentration in the indium-oxides. Unfortunately,
conductance relaxation in low n systems appears to be rather fast (see
\cite{6,10} for a discussion of this issue), in which case the qualitative
change of behavior in $\Delta$G(t)~excited by low vs. high frequencies may be
hard to observe.

The phenomenology associated with the MW-induced excess-conductivity, in
particular, the lack of slow relaxation and the sublinear dependence on power,
puts constraints on the underlying mechanism. Detailed examination of the
glassy features in the presence of MW radiation rules-out heating as the
reason of the effect. A promising direction is a mechanism based on a model
recently proposed by M\"{u}ller and Ioffe \cite{16}. In this scenario the MW
drives collective electronic-modes that act as an extra energy-source thus
enhancing hopping processes. This scenario is currently being explored.

Discussions with A. Amir, A. Efros, Y. Imry, and M. Pollak are gratefully
acknowledged. This research was supported by a grant administered by the US
Israel Binational Science Foundation and by the Israeli Foundation for
Sciences and Humanities.


\begin{thebibliography}{99}                                                                                               %


\bibitem {1}J. H. Davies et al, Phys. Rev. Lett, \textbf{49}, 758 (1982); M.
Gr\"{u}newald et al, J. Phys. C, \textbf{15}, L1153 (1982); M. Pollak and M.
Ortu\~{n}o, Sol. Energy Mater., \textbf{8}, 81 (1982); M. Pollak, Phil. Mag.
B\textbf{ 50}, 265 (1984); G. Vignale, Phys. Rev. B\textbf{ 36}, 8192 (1987);
M. M\"{u}ller and L. B. Ioffe, Phys. Rev. Lett. \textbf{93}, 256403 (2004); C.
C. Yu, Phys. Rev. Lett., \textbf{82}, 4074 (1999); Vikas Malik and Deepak
Kumar, Phys. Rev. B \textbf{69}, 153103 (2004); D. R. Grempel, Europhys.
Lett., \textbf{66,} 854 (2004); Eran Lebanon, and Markus M\"{u}ller, Phys.
Rev. B\textbf{\ 72}, 174202 (2005).

\bibitem {2}M. Ben Chorin et al, Phys. Rev. B\textbf{ 48}, 15025 (1993).

\bibitem {3}M. Ben-Chorin et al, Phys. Rev. B \textbf{44}, 3420 (1991); G.
Martinez-Arizala et al, Phys. Rev. Lett., \textbf{78}, 1130 (1997); Z.
Ovadyahu and M. Pollak, Phys. Rev. Lett., \textbf{79}, 459 (1997); T. Grenet,
Eur. Phys. J, \textbf{32}, 275 (2003).

\bibitem {4}A. Vaknin et al, Phys. Rev. B \textbf{61}, 6692 (2000); Ariel Amir
et al, Phys. Rev. B \textbf{77}, 165207 (2008).

\bibitem {5}A. Vaknin et al, Phys. Rev. B\textbf{ 65}, 134208 (2002).

\bibitem {6}Z. Ovadyahu, Phys. Rev. B \textbf{78}, 195120 (2008).

\bibitem {7}M. Pollak, Discuss. Faraday Soc. \textbf{50}, 13 (1970); M. L.
Knotek and M. Pollak, J. Non-Cryst. Solids, 505, \textbf{8-10} (1972); A. L.
Efros and B. I. Shklovskii, J. Phys. C \textbf{8}, L49 (1975).

\bibitem {8}S. D. Baranovski et al, J. Phys. C: Solid State Phys.,
\textbf{12}, 1023 (1979).

\bibitem {9}R. G. Palmer at al; M. M\"{u}ller and S. Pankov, Phys. Rev. B
\textbf{75}, 144201 (2007); M. M\"{u}ller and S. Pankov, Phys. Rev. B
\textbf{75}, 144201 (2007).

\bibitem {10}Z. Ovadyahu, Phys. Rev. B \textbf{73}, 214208 (2006); Phys. Rev.
Lett., \textbf{99}, 226603 (2007).

\bibitem {11}A. Vaknin et al, Europhys. Letters \textbf{42}, 307 (1998).

\bibitem {12}V. Orlyanchik, and Z. Ovadyahu, Phys. Rev. Lett., \textbf{92},
066801 (2004).

\bibitem {13}M. E. Gershenson et al, Phys. Rev. Lett. \textbf{85}, 1718 (2000).

\bibitem {14}A. Vaknin et al, Phys. Rev. Lett., \textbf{81}, 669 (1998).

\bibitem {15}E. Helgren et al., Phys. Rev. B \textbf{69}, 014201 (2004); M.
Hering et al., Physica B \textbf{359-361}, 1469 (2005).

\bibitem {16}Markus M\"{u}ller, and Lev B. Ioffe, cond-mat/arXiv:0711.2668
\end{thebibliography}
\end{document}